\documentclass[a4paper]{JHEP3}
\usepackage[centertags]{amsmath}
\usepackage{amssymb}
\usepackage{fixltx2e}
\MakeRobust{\eqref}
\usepackage{graphicx}
\usepackage{cite}
\usepackage{subfigure}
\preprint{TIFR/TH/14-20\\}

\newcommand{\Tr}{\text{Tr}}

\def \beal#1 {\begin{align}#1\end{align}}

\def\Tr{\mathrm{Tr}}

\newcommand{\wt}{\widetilde}

\def\={\stackrel{\bullet}{=}}

\def\Tr{\mathrm{Tr}}

\def\[{\left[}
\def\]{\right]}
\def\({\left(}
\def\){\right)}

\def \be {\begin{equation}}
\def \ee {\end{equation}}
\def \bea {\begin{eqnarray}}
\def \eea {\end{eqnarray}}
\def \beal#1 {\begin{align}#1\end{align}}

\def \be {\begin{equation}}
\def \ee {\end{equation}}
\def \bea {\begin{eqnarray}}
\def \eea {\end{eqnarray}}
\def \beal#1 {\begin{align}#1\end{align}}

\usepackage{enumerate}

\title{Poles in the S-Matrix of Relativistic Chern-Simons Matter theories 
from Quantum Mechanics}

\author{
Yogesh Dandekar$^{a),1}$, 
Mangesh Mandlik$^{a),2}$, 
Shiraz Minwalla$^{a),3}$
\\
$^{a)}$Department of Theoretical Physics, Tata Institute of Fundamental Research,
Homi Bhabha Road, Mumbai 400005, India\\
{\small \tt E-mail: $^1$yogesh@theory.tifr.res.in, ${}^2$mangesh@theory.tifr.res.in,  ${}^3$minwalla@theory.tifr.res.in}
}

\abstract{An all orders formula for the S-matrix for 2 $\rightarrow$
2 scattering in large N Chern-Simons theory coupled to a fundamental
scalar has recently been conjectured. We find a scaling limit of the theory
in which the pole in this S-matrix is near threshold.
We argue that the theory must be well described by non-relativistic quantum
mechanics in this limit, and determine the relevant Schroedinger equation.
We demonstrate that the S-matrix obtained from this Schroedinger equation
agrees perfectly with this scaling limit of the relativistic S-matrix; in
particular the pole structures match exactly.  We view this matching
as a nontrivial consistency check of the conjectured field theory S-matrix.   
}

\begin{document}

\section{Introduction, analysis and conclusions}

Pure Chern-Simons theories in three dimensions are topological and 
have no local degrees of freedom; these theories have been studied 
in great detail over the last three decades starting with the classic work 
of Witten \cite{Witten:1988hf}.

Relativistic Chern-Simons theories minimally coupled to charged matter fields 
have been studied much less than pure Chern-Simons theories.
\footnote{Non-relativistic versions of these theories 
describe interactions of anyonic particles, and have been studied in various 
guises over the last few decades, starting with the work of Aharonov and 
Bohm \cite{Aharonov:1959fk} (see e.g. the book \cite{wilczek1990fractional} for 
a review and \cite{Jackiw:1989qp,Bak:1994dj,Bak:1994zz,AmelinoCamelia:1994we,Park:2012yw} for more recent work).} 
These theories are non topological, have local degrees of freedom and are not 
generically solvable. However it has recently been realized that 
$U(N)$ level $\kappa$ Chern-Simons theories coupled to fundamental matter are 
effectively solvable in the 't Hooft large $N$ limit, $N \to \infty$ and 
$\kappa \to \infty$ with $\frac{N}{\kappa}=\lambda$ held fixed. \footnote{$\kappa$ is the level of the Chern-Simons term in the bulk Lagrangian in the dimensional reduction scheme. It is defined as $\kappa = sgn(k)(|k|+N)$, where $|k|$ is the level of the WZW theory dual to the pure Chern-Simons theory. Please note that $\kappa$ used in this paper = $k$ used in \cite{Jain:2014nza}} Three point correlators 
and the thermal partition function have been computed for
such theories to all orders in the 't Hooft coupling $\lambda$; the 
results of these computations have motivated the conjecture for non 
supersymmetric strong weak coupling bosonization type dualities between 
pairs of these theories, and also for conjectured Vasiliev bulk duals 
for these theories.
\footnote{In a parallel development there has been an 
explosion of exact 
results for protected quantities (indices, partition functions etc.) for 
supersymmetric Chern-Simons theories, obtained using the methods of 
supersymmetric localization. As it is possible to construct supersymmetric 
theories with matter in the fundamental representation, it should be possible 
- and would be fascinating - to make connections between these as yet 
distinct streams of work.}
\cite{Aharony:2011jz,Giombi:2011kc,Maldacena:2011jn,Maldacena:2012sf,Chang:2012kt,
Aharony:2012nh,Jain:2012qi,Yokoyama:2012fa,
GurAri:2012is,Aharony:2012ns,Jain:2013py,Takimi:2013zca,Jain:2013gza,Frishman:2013dvg,Yokoyama:2013pxa}

The recent paper \cite{Jain:2014nza} has initiated the study of the 
S-matrix in fundamental matter Chern-Simons theories to all orders in the 
't Hooft coupling. In particular the authors of \cite{Jain:2014nza} 
have presented a detailed study of $2 \to 2$ scattering in the most
general renormalizable theory of a single fundamental scalar interacting 
with a $U(N_B)$ Chern-Simons gauge field
\begin{equation}\label{lag}
S  = \int d^3 x  \biggl[i \varepsilon^{\mu\nu\rho} \frac{\kappa_{B}}{4 \pi}
\Tr( A_\mu\partial_\nu A_\rho -\frac{2 i}{3}  A_\mu A_\nu A_\rho)
+  D_\mu \bar \phi  D^\mu\phi 
+m_B^2 \bar\phi \phi+\frac{1}{2 N_B} b_4(\bar{\phi}\phi)^2\biggl],
\end{equation}
to all orders in $\lambda_B = \frac{N_B}{\kappa_B}$.

The theory \eqref{lag} has elementary quanta that transform in either 
the fundamental or the antifundamental representations of 
$U(N_B)$. Following the terminology of \cite{Jain:2014nza}, we refer to quanta in the 
fundamental representation as particles, and quanta in the antifundamental 
representation as antiparticles. The authors of \cite{Jain:2014nza} 
were able to explicitly compute the particle - particle scattering matrix 
together with the particle - antiparticle scattering matrix in the 
channel corresponding to adjoint exchange. They also presented the following 
conjectured formula for the particle - antiparticle S-matrix in the 
channel corresponding to singlet exchange:
\begin{equation}\label{smatrix}
\begin{split}
T_S(\sqrt{s},\theta) &= 8\pi i \sqrt{s}(1-\cos(\pi \lambda_{B}))\delta(\theta)+ 4 i \sqrt{s}\sin(\pi \lambda_{B}) {\rm Pv} \left(\cot\left(\frac{\theta}{2}\right)\right)\\
 &+ 4 \sqrt{s}\sin(\pi |\lambda_B|)  
\left( \frac{ \left(4 \pi  |\lambda_B| \sqrt{s} +\wt b_4 \right)
  +  e^{i\pi|\lambda_B|} \left(-4 \pi  |\lambda_B| \sqrt{s} + \wt b_4 \right)  \left(  \frac{ \frac{1}{2} + \frac{c_B}{\sqrt{s} } }
{\frac{1}{2} - \frac{c_B}{\sqrt{s}} } \right)^{|\lambda_B|}
 } {\left(4 \pi  |\lambda_B| \sqrt{s}  + \wt b_4 \right) 
 -  e^{i\pi|\lambda_B|} \left(-4 \pi  |\lambda_B| \sqrt{s} + \wt b_4 \right)    \left(  \frac{ \frac{1}{2} + \frac{c_B}{\sqrt{s} } }
{\frac{1}{2} - \frac{c_B}{\sqrt{s}} } \right)^{|\lambda_B|} 
} \right),
\end{split}
\end{equation}
where
\begin{equation}\label{where}
\begin{split}
c_B &= {\rm pole~mass~of~the~single~scalar~excitation},\\
\sqrt{s} &= {\rm centre~of~mass~energy},\\
\theta &= {\rm angle~of~scattering},\\
\wt b_4 &= 2\pi\lambda_B^2c_B-b_4.
\end{split}
\end{equation}

As explained in \cite{Jain:2014nza}, the S-matrix \eqref{smatrix} 
does {\it not} agree with the simple analytic continuation of the 
particle - particle S-matrix. Instead, the nonsingular part of 
\eqref{smatrix} is given by the analytic continuation of the particle - particle S-matrix rescaled by 
the factor $\frac{\sin \left( \pi \lambda_B \right)}{\pi \lambda_B}$. In other words  
the correctness of the conjectured S-matrix \eqref{smatrix} requires 
an intriguing  modification of the usual text book rules of crossing 
symmetry in the case of matter Chern-Simons theories. As with any conjecture
that challenges accepted wisdom, the formula \eqref{smatrix} should be 
subjected to stringent checks. In this note we confront the conjecture 
of \cite{Jain:2014nza} with a  nontrivial consistency check and find 
that it passes the test, as we now describe.
 
The S-matrix \eqref{smatrix} has a pole for 
$\widetilde{b}_4 \geq \widetilde{b}_4^{crit}=8\pi c_B |\lambda_B| $ indicating the existence 
of a particle - antiparticle bound state in the singlet channel at these 
values of parameters.
\footnote{$~~b_4$ is always negative when bound states exist,  
so it possible that \eqref{lag} is non perturbatively unstable in this range of parameters. While the study of the nonperturbative stability of \eqref{lag} is an interesting question (one that can presumably be settled by the evaluation of the all orders  effective action for $\phi$),  it is irrelevant for the perturbative considerations of this note, and will not be studied in this paper.}
As $\widetilde{b}_4$ approaches  $\widetilde{b}_4^{crit}$ from above,  the mass of the bound state approaches $2c_B$. In other words, if we set 
$\widetilde{b}_4=\widetilde{b}_4^{crit} +\delta b_4$, the binding energy 
$E_B$ is small at small $\delta b_4$ 
(it turns out $E_B \sim (\delta b_4)^{1/|\lambda_B|}$)
\footnote{More precisely, at lowest nontrivial order in $\delta b_4$  
\begin{equation}\label{lno}
\frac{E_B}{4c_B}= 
\left( \frac{\delta b_4}{16 \pi |\lambda_B| c_B}\right)^\frac{1}{|\lambda_B|}.
\end{equation} }
 and vanishes when 
$\delta b_4=0$. 

Motivated by this observation, in this note we focus on the field theory 
\eqref{lag} in a sector containing a singlet particle - antiparticle pair 
in a particular scaling limit we call the `near threshold limit'. 
This limit is defined by scaling $\delta b_4$ to zero while simultaneously 
scaling $\sqrt{s}-2c_B$ to zero like $(\delta b_4)^{1/|\lambda_B|}$. In this limit 
the particles are non-relativistic and we may set $\sqrt{s}-2 c_B= \frac{k^2}{c_B}$.\footnote{Note that our definition of the near threshold 
limit does not constrain $\sqrt{s}-2c_B$ to take a particular sign. This quantity 
is negative in the study of bound states, and positive in the study of 
scattering.}
In our scaling limit
\begin{equation}\label{cnrl}
\frac{\delta b_4}{c_B} \to 0, ~~~ \frac{k}{c_B} \to 0, ~~~
\frac{k}{c_B} \left( \frac{c_B}{\delta b_4} \right)^{\frac{1}{2 |\lambda_B|} } =~{\rm fixed}.
\end{equation}

Like any non-relativistic limit, our limit 
focuses attention on a sector of the theory in which kinetic energies of the 
particle and antiparticle are small compared to rest masses. 
In this limit our system must admit an effective description in terms of 
the non-relativistic quantum mechanics of two particles interacting via 
Chern-Simons gauge boson exchange, plus a contact interaction. We will 
now describe this quantum mechanical system in more detail, following
Amelino-Camelia and Bak \cite{AmelinoCamelia:1994we}.

It is well known (see, for instance, \cite{Bak:1994dj,Bak:1994zz}) that the entire effect of the Chern-Simons
interactions between non-relativistic particles is to implement anyonic 
statistics for the particles. This happens because the Chern-Simons equation 
of motion forces each particle to trap a unit of flux; the other particle
picks up a phase when circumnavigating this flux. The magnitude of the phase depends 
on the coupling colour factors: when the colour factors of the two particles (which transform 
in representations $R_1$ and $R_2$ respectively) Clebsch-Gordon couple into representation $R_m$ 
it turns out that the magnitude of the phase is given by \cite{Bak:1994dj} 
\begin{equation}\label{phase}
\nu_m=\frac{c_2(R_m) -c_2(R_1) -c_2(R_2)}{\kappa},
\end{equation}
where $c_2(R)$ is the quadratic Casimir of the representation $R$. 

The effect of this phase is most 
simply described when we change variables to work with the centre of 
mass and relative degrees of freedom of the particle - antiparticle 
system. The centre of mass motion is free, and is ignored in 
what follows. In terms of relative coordinates, in the gauge singlet sector 
(i.e. $c_2(R_m)=0$), the entire effect of the 
Chern-Simons coupled gauge field is implemented by inserting a point 
like solenoid of integrated flux $-2 \pi \lambda_B$ at the origin of the two 
dimensional plane. The quantum mechanical description of this system 
is given by a non-relativistic Schroedinger equation \eqref{sedf} below for a particle of 
effective mass $\frac{c_B}{2}$ and of effective $U(1)$ charge unity, 
minimally coupled to a $U(1)$ gauge field corresponding to this point like solenoid
(see the section (2.6) of \cite{Jain:2014nza} and references therein). In other words, the time
independent Schroedinger equation for our system at energy $E=\sqrt{s}-2c_B
=\frac{k^2}{c_B}$ is given by \begin{equation}\label{sedf}
 \begin{split}
 -D_iD^i {\psi} & = k^2 \psi, \\
 D_i & = \nabla_i + i A_i, \\
 A_i& = \nu \frac{ \epsilon_{i j} x^j}{x^2},
 \end{split}
\end{equation}
where, in the singlet sector, (as in (2.47) of \cite{Jain:2014nza})
\begin{equation}
 \nu = -\lambda_B.
\end{equation}

It turns out that the point like interaction between the particle and the 
antiparticle imposes modified boundary conditions for this effective 
Schroedinger wave function at origin \cite{AmelinoCamelia:1994we,Kim:1996rz}(see the Appendix \ref{reb} for an intuitive explanation).  
As explained in  \cite{AmelinoCamelia:1994we,Kim:1996rz} there exists a one parameter 
set of consistent and self-adjoint boundary conditions for the wave function 
at the origin. These boundary conditions are specified 
as follows.  Let 
\begin{equation}\label{pft}
\psi({\vec r})= \sum_m e^{i m \theta} \psi_m(r).
\end{equation}
The functions $\psi_m(r)$ for $m \neq 0$ are required, as usual to vanish 
at $r=0$. For $m=0$, on the other hand, we require that 
\begin{equation} \label{pbc}
\psi_0(r) \propto \left( r^{|\lambda_B|} + \frac{ w R^{2 |\lambda_B|} }
{r^{|\lambda_B| } } \right),
\end{equation}
where $R$ is a reference length scale and $w$ is the self-adjoint extension parameter as introduced in \cite{AmelinoCamelia:1994we}.

In other words $\psi_0$ is not forced to vanish at the origin but has 
a component that blows up. We refer to \eqref{pbc} as the Amelino-Camelia-Bak boundary conditions.  

The modified boundary conditions \eqref{pbc} are labeled by the single 
dimensionful parameter $ wR^{2 |\lambda_B|}$. It follows from dimensional 
analysis that the effect of this parameter on any process with 
characteristic momentum scale $k$ (like the scattering of particles 
with momentum $k$) is proportional to $ w (Rk)^{2 |\lambda_B|}$. 
As $ w (Rk)^{2 |\lambda_B|} \rightarrow 0$ the boundary conditions above
effectively reduce to the `usual' Aharonov-Bohm boundary conditions; the boundary 
conditions that force $\psi_0$ to vanish at the origin.

In summary, the low energy effective description of the particle - 
antiparticle system in the near threshold 
limit is given by the quantum mechanics of a single non-relativistic 
particle propagating in two dimensions. The wave function of this 
particle obeys the Schroedinger equation \eqref{sedf} and the boundary 
conditions \eqref{pbc}. The boundary condition parameter $w R^{2 |\lambda_B|}$ in
\eqref{pbc} is an as yet unknown function of $\delta b_4$.

It follows from the discussion above that the S-matrix 
\eqref{smatrix} must reduce in the near threshold limit, 
to the S-matrix computed by solving \eqref{sedf} subject to the Amelino-Camelia- Bak boundary conditions. This expectation is a nontrivial consistency 
check of the conjecture \eqref{smatrix}, which we now proceed to verify. 

The near threshold limit of the S-matrix \eqref{smatrix} is 
easily determined. As above we set
\begin{equation} \label{setting}
\sqrt{s}= 2 c_B  + \frac{k^2}{c_B}.
\end{equation}
In the limit \eqref{cnrl}, the second line of \eqref{smatrix} reduces to 
\begin{equation*}\label{jred}
8 c_B 
|\sin(\pi\lambda_B)|\frac{1+e^{i\pi|\lambda_B|} \left[ \frac{\delta b_4 \left( \frac{2 c_B}{k} \right)^{2 |\lambda_B|}}{16 \pi |\lambda_B| c_B
 }\right]}
{1-e^{i\pi|\lambda_B|} \left[\frac{\delta b_4 \left( \frac{2 c_B}{k} \right)^{2 |\lambda_B|}}{16 \pi |\lambda_B| c_B
 }\right]},
\end{equation*}
so that the S-matrix \eqref{smatrix} reduces to 
\begin{equation}\label{smnr}
\begin{split}
T_S(\sqrt{s},\theta) &= -16 \pi i c_B(\cos(\pi \lambda_{B})-1)\delta(\theta)+ 8 i c_B \sin(\pi \lambda_{B}){\rm Pv}\left( \cot\left(\frac{\theta}{2}\right)\right)\\
&+ 8 c_B 
|\sin(\pi\lambda_B)|\frac{1+e^{i\pi|\lambda_B|} \frac{A_{R}}{k^{2|\lambda_B|}} }
{1-e^{i\pi|\lambda_B|} \frac{A_{R}}{k^{2|\lambda_B|}} },\\
&A_R=  \left[ \frac{\delta b_4 \left( 2 c_B \right)^{2 |\lambda_B|}}{16 \pi |\lambda_B| c_B
 }\right].
\end{split}
\end{equation}

On the other hand the S-matrix obtained by solving the Schroedinger 
equation \eqref{sedf} subject to the boundary conditions \eqref{pbc} 
has already been determined in  \cite{AmelinoCamelia:1994we} 
and we rederive it in the Appendix \ref{abap}. \footnote{More precisely in the Appendix \ref{abap} we show 
that the Schroedinger equation described above has a scattering 
solution that takes the form
\begin{equation}\label{scatsol} \begin{split}
\psi({\vec r})& = e^{i k x} + \zeta({\vec r} ),\\
\zeta({\vec r})& =\frac{e^{- \frac{i \pi}{4}} e^{i k r} h(\theta)}{
\sqrt{2 \pi k r} } + {\cal O} \left( \frac{1}{r^{\frac{3}{2}}} \right),\\
h(\theta)&= 2 \pi \left( \cos \left( \pi \lambda_B \right) - 1 \right) 
\delta(\theta) - \sin (\pi \lambda_B) {\rm Pv} \left( \cot \frac{\theta}{2} \right)
+ i |\sin \left(\pi \lambda_B\right)| \frac{1+e^{i\pi|\lambda_B|} \left[\frac{-1}{w}\left(\frac{2}{kR}\right)^{2|\lambda_B|}\frac{\Gamma(1+|\lambda_B|)}{\Gamma(1-|\lambda_B|)}\right]}{1-e^{i\pi|\lambda_B|} \left[ \frac{-1}{w}\left(\frac{2}{kR}\right)^{2|\lambda_B|}\frac{\Gamma(1+|\lambda_B|)}{\Gamma(1-|\lambda_B|)}\right]}.
\end{split}
\end{equation}
The non-relativistic limit of the usual invariant scattering amplitude is given by
\begin{equation} \label{sa}
T_{NR}= -8 i c_B h(\theta).
\end{equation}}
It turns out that 
\begin{equation}\label{bcsm}\begin{split}
T_{NR}&= -16 \pi i c_B \left( \cos \left( \pi \lambda_B \right) - 1 \right) 
\delta(\theta) +8 i c_B \sin (\pi \lambda_B) {\rm Pv} \left( \cot \frac{\theta}{2}\right) \\
& + 8 c_B |\sin \pi \lambda_B| \frac{1+e^{i\pi|\lambda_B|} \frac{A_{NR}}{k^{2|\lambda_B|}}}
{1-e^{i\pi|\lambda_B|} \frac{A_{NR}}{k^{2|\lambda_B|}} },\\
&A_{NR}= \frac{-1}{w}\left(\frac{2}{R}\right)^{2|\lambda_B|}\frac{\Gamma(1+|\lambda_B|)}{\Gamma(1-|\lambda_B|)}.
\end{split}
\end{equation}
The $S$-matrices \eqref{smnr} and \eqref{bcsm} are identical in structure. They agree in all details 
provided we identify
\begin{equation} \label{ident} 
-w \left( c_B R \right)^{2 |\lambda_B|}= \frac{c_B}{\delta b_4} 
 \left( 16 \pi |\lambda_B| 
\frac{\Gamma(1+ |\lambda_B|)}{\Gamma(1-|\lambda_B|)} \right).
\end{equation}
\eqref{ident} determines the hitherto unknown dependence of the boundary 
condition parameter $w R^{2 |\lambda_B|}$ as a function of $\delta b_4$. 

In summary, in the near threshold limit, 
the S-matrix \eqref{smatrix} agrees perfectly with the 
S-matrix computed from the Schroedinger equation 
\eqref{sedf} subject to the boundary conditions 
\begin{equation} \label{pbcr}
\psi_0(r) \propto \left(  r^{|\lambda_B|}  
- \frac{  \frac{c_B}{\delta b_4} 
 \left( 16 \pi |\lambda_B| 
\frac{\Gamma(1+ |\lambda_B|)}{\Gamma(1-|\lambda_B|)} \right)}
{\left(r c_B^2 \right) ^{|\lambda_B| } } \right).
\end{equation}

As we have emphasized above, however, the 
effect of the modified boundary conditions on 
a process at momentum scale $k$ is measured by $w (Rk)^{2 |\lambda_B|}$. It follows from 
\eqref{ident} that in the current situation, the effect of the modified boundary 
conditions on a process at momentum scale $k$ is measured by   
 \begin{equation}\label{tm} M= \frac{c_B}{\delta b_4} 
\left( \frac{k}{c_B} \right) ^{ 2|\lambda_B| }
 \left( 16 \pi |\lambda_B| 
\frac{\Gamma(1+ |\lambda_B|)}{\Gamma(1-|\lambda_B|)} \right).
\end{equation}
Note that $M$ is held fixed in the near threshold 
scaling limit \eqref{cnrl}. The modified boundary condition can be ignored 
when $M \to 0$. $M$ tends to zero in, for instance, the usual 
non-relativistic limit  (where $k$ is scaled to zero with all other parameters like 
$\delta b_4$ held fixed). Consequently $w R^{2 |\lambda_B|}$ is effectively 
zero in the quantum mechanical description of the usual non-relativistic limit, 
explaining why  \eqref{smatrix} reduces to the $w=0$ Aharonov-Bohm-Ruijsenaars  \cite{Aharonov:1959fk,Ruijsenaars:1981fp} S matrix in 
this limit, as noted in \cite{Jain:2014nza}.

The agreement of the S-matrix \eqref{smatrix} (and in particular of its poles) 
with \eqref{bcsm} in the near threshold limit immediately 
demonstrates that the spectrum of near threshold bound states of the 
singlet particle - antiparticle sector of \eqref{lag} agrees with 
the spectrum of bound states of the Schroedinger equation \eqref{sedf} 
subject to the boundary conditions \eqref{pbcr}. 

The scattering matrices $T_S$ and $T_{NR}$ are quite involved functions of $k$ and $\lambda_B$; 
for this reason we view the matching of these two functions in the appropriate limit as a rather 
nontrivial test of the conjectured S matrix \eqref{smatrix}. Note that $T_S$  would not have 
matched with $T_{NR}$ without the the additional factor $\frac{\sin (\pi \lambda_B)}
{\pi \lambda_B}$ invoked in \cite{Jain:2014nza}. As a consequence the results of this note provide
indirect support to the modified crossing symmetry properties for the S matrix of matter Chern-Simons theories
conjectured in \cite{Jain:2014nza}.

In this paper we have argued that the S matrix \eqref{smatrix} may be derived from a Schroedinger equation 
in a particular scaling limit. Perhaps it is possible to derive the full relativistic formula \eqref{smatrix}
from the solution to an appropriate Schroedinger equation in lightcone slicing; we leave the further 
investigations of this issue to future work.

\acknowledgments
We would like to thank K. Inbasekar, S. Jain, J. Maldacena, D. Son, T. Takimi, S. Wadia and S. Yokoyama
for useful discussions and S. Jain for comments on the manuscript. S.M. would like to thank the University of Chicago and especially the 
Institute for Advanced Study for hospitality when this work was initiated. We would also like to 
acknowledge our debt to the people of India for their steady and generous support to research in the basic 
sciences.

\appendix

\section{The quantum mechanics of anyons with point like interactions}

In the main text we have followed \cite{AmelinoCamelia:1994we,Kim:1996rz} to assert that point like interactions 
between anyons effectively impose modified local boundary conditions on the 
Schroedinger equation in the relative coordinates. This assertion may 
appear unfamiliar as contact interactions usually lead to delta function 
potentials for relative coordinates. In fact these viewpoints are equivalent. 
In the subsections \ref{rec} and \ref{reb} below we demonstrate that 
the correct treatment of the two dimensional delta function at $\lambda_B=0$ does, in fact, 
effectively modify the boundary conditions at the origin and has no other 
effect. Moreover the boundary conditions so obtained agree with $\lambda_B \to 0$ 
limit of the boundary conditions \eqref{pbc}.

In subsection \ref{abap} we proceed to rederive the scattering amplitude for the Schroedinger 
equation \eqref{sedf} subject to the boundary conditions \eqref{pbc}; our results agree 
with those of \cite{AmelinoCamelia:1994we}.

\subsection{Quantum mechanics with a two dimensional delta function}

\subsubsection{Renormalization of the coupling constant} \label{rec}

In this section we review the dynamics of the quantum mechanical system 
governed by the two dimensional Schroedinger equation
\begin{equation}\label{sedfr}
 -\frac{\nabla^2}{2m} {\psi({\vec x})} + V({\vec x}) \psi({\vec x}) = \frac{k^2}{2m} \psi({\vec x}),
\end{equation}
where $V({\vec x})$ is taken to be proportional to a  suitably renormalized version of the 
attractive two dimensional $\delta$ function. 
This system has been studied in 
great detail in several papers (see e.g. \cite{Jackiw:1991je}); we review the principal results. 

Let 
\begin{equation}\label{momspace}
 \psi({\vec x})= \int \frac{d^2 k}{(2 \pi)^2} e^{i {\vec k}.{\vec x}} {\widetilde \psi}({\vec k}).
\end{equation}
The time independent solution of \eqref{sedfr} that describes the scattering of an incoming 
particle with momentum ${\vec k}$ off an arbitrary potential $V(x)$ 
 is given by the solution to the Lippmann-Schwinger equation 
\begin{equation}\label{LS}
{\widetilde \psi }({\vec p})= (2\pi)^2 \delta^2({\vec p}-{\vec k}) +  2m\int \frac{d^2 q}{(2 \pi)^2} \frac{ {\widetilde V}({\vec q}) {\widetilde \psi}({\vec p}-{\vec q}) }{k^2-p^2 +i \epsilon}.
\end{equation}
Let $V(x)= - g \delta^2({\vec x})$ so that its Fourier transform is given by 
${\widetilde V}({\vec k})= -g$. Plugging into \eqref{LS} 
we find 
\begin{equation} \label{gapf}
{\widetilde \psi}({\vec p})=  (2\pi)^2 \delta^2({\vec p}-{\vec k}) - \frac{2mg A ({\vec k })}{k^2-p^2 + i \epsilon},
\end{equation}
where 
\begin{equation}\label{gape}
A({\vec k})= \frac{1}{1-2mg \int \frac{d^2 p}{(2 \pi)^2}  \frac{1}{p^2 - k^2 - i \epsilon}}.
\end{equation}
The integral in \eqref{gape} diverges logarithmically. Evaluating the integral with a 
cut off $\Lambda$ we have
\begin{equation}\label{grape}
A({\vec k})= \frac{1}{1- \frac{mg}{2 \pi} \ln  \left( \frac{\Lambda^2}{-k^2} \right)}.
\end{equation}

The function $A({\vec k})$ is proportional to the scattering amplitude of our 
quantum mechanical system. In order to define a sensible 
scattering problem we must regulate and renormalize \eqref{grape} 
by choosing the coupling constant $g$ to scale to zero logarithmically with the cut off $\Lambda$. 
We choose $g(\Lambda)$ so that 
\begin{equation}\label{rg}
 \frac{1}{g(\Lambda)}=  \frac{1}{g_R(\mu)}+ \frac{m}{2\pi}\ln\left(\frac{\Lambda^2}{\mu^2}\right),
\end{equation}
where the renormalized coupling $g_R(\mu)$ is held fixed as $\Lambda$ is taken to infinity. $g_R(\mu)$ is, of course, 
a function of the renormalization scale $\mu$. \eqref{grape} now takes the form 
\footnote{As an application 
notice that the scattering amplitude \eqref{samp} has a pole at 
\begin{equation}
 k^2 = -\mu^2e^{-\frac{2\pi}{mg_R}},
\end{equation}
implying that our renormalized $\delta$ function potential quantum mechanics 
has a single bound state with binding energy
\begin{equation}\label{beqm}
 E=-\frac{\mu^2}{2m}e^{-\frac{2\pi}{mg_R}}.
\end{equation} }
\begin{equation}\label{samp} 
A({\vec k})=\frac{1}{1- \frac{mg_R}{2 \pi} \ln  \left( \frac{\mu^2}{-k^2}\right)}.
\end{equation}

\subsubsection{Description in terms of modified boundary conditions} \label{reb}

We will now find an alternative effective description of the renormalized two dimensional 
delta function in terms of modified boundary conditions at $r=0$. 
For this purpose it will prove convenient to work in position rather 
than momentum space. For this reason we regulate the $\delta$ function 
potential as the `circular square well'
\begin{equation}\label{cirsw}
\begin{split}
 V(r)&=- \frac{g}{\pi r_0^2} \ \ ;\ r<r_0 ,\\
&=0  ~~~~~~\ \ ;\ r>r_0.
\end{split}
\end{equation}

Let us now study rotationally invariant solutions of the two dimensional 
Schroedinger equation with the potential \eqref{cirsw}. \footnote{Only rotationally invariant 
solutions are affected by the potential \eqref{cirsw} in the limit $r_0 \to 0$, as the wave function
at nonzero angular momentum dies rapidly at small $r$ due to the angular momentum 
barrier.} Clearly the most general regular (at $r=0$) 
solution to the Schroedinger equation takes the form 
\begin{equation}
\begin{split}
 aJ_0(lr)\ ;\ r<r_0 ,\\
 cJ_0(k r)+dY_0(k r) \ ;\ r>r_0,
\end{split}
\end{equation}
where,
\begin{equation}
 l^2=2m\left(\frac{g}{\pi r_0^2}+E\right)\ \ ,\ \ k^2=2mE.
\end{equation}
The requirement of continuity of the wave function and its first derivative 
across $r=r_0$ determines $d$ and $c$ in terms of $a$. In the small $r_0$ 
limit it is easily verified that 
\begin{equation}\label{dcbare}
 \frac{d}{c}=\frac{-1}{\frac{2}{mg}+\frac{2}{\pi}\left[\gamma+\ln\left(\frac{k r_0}{2}\right)\right]},
\end{equation}
where, $\gamma$ is Euler-Mascheroni constant. 

As in the previous section \eqref{dcbare} does not have a well defined 
$r_0 \to 0$ limit. In order that the LHS of \eqref{dcbare} is well defined 
as $r_0 \to 0$ we must choose $g$ to be a function of $r_0$ and take $g$ to zero as $r_0$ is 
scaled to zero, keeping $g_R$ fixed where 
\begin{equation} \label{rgm}
 \frac{1}{g_R(\mu)}=\frac{1}{g(r_0)}+\frac{m}{\pi}\left[\ln \left(\frac{r_0\mu}{2}\right)+\gamma\right].
\end{equation}
Note that\eqref{rgm} agrees exactly with \eqref{rg} under the replacement 
$\frac{\mu r_0 e^{\gamma}}{2}\rightarrow\frac{\mu}{\Lambda}$.

Implementing this limit we find
\begin{equation}\label{bcwonu}
 \frac{d}{c}=\frac{-1}{\frac{2}{mg_R}+\frac{2}{\pi}\ln \left(\frac{k}{\mu}\right)}.
\end{equation}
It follows that the Schroedinger problem with a delta function potential with 
renormalized strength $g_R$ is equivalent to the free Schroedinger 
equation subject to the $r \to 0$ boundary condition   
\begin{equation}\label{mbc}
\psi_0(r) \propto  \left[ \left( -\frac{2}{m g_R} - \frac{2}{\pi} \ln \frac{k}{\mu} 
\right) J_0\left(kr\right) + Y_0(kr) \right].
\end{equation}
Using the small argument expansions
\begin{equation}\label{JYrz}
J_0(kr)=1 + {\cal O}\left((kr)^2\right), ~~~Y_0(kr)=\frac{2}{\pi} \ln \left(\frac{kr}{2}\right) + 
2 \frac{\gamma}{\pi} +{\cal O}\left((kr)^2 \ln (kr)\right),
\end{equation} 
we see that the $k$ dependence cancels from \eqref{mbc} and the boundary 
condition on $\psi(r)$ takes the local form 
\begin{equation}\label{mbcc}
\psi_0(r) \propto \left[ \left( -\frac{2}{m g_R} + 2 \frac{\gamma}{\pi} \right) 
+ \frac{2}{\pi} \ln \left( \frac{\mu r}{2}\right) +\mathcal{O}\left(r^2 \ln r\right)\right].
\end{equation}

In summary, the Schroedinger equation in the presence of a renormalized 
$\delta$ function potential is exactly equivalent to the free 
Schroedinger equation subject to the local boundary conditions 
\eqref{mbcc} at the origin.

It is easily verified that the boundary conditions \eqref{mbcc} are obtained as a limit of 
the Amelino-Camelia-Bak boundary conditions \eqref{pbc} if we set 
$$w= -1+ |\lambda_B| \left( -\frac{2 \pi}{m g_R} + 2 \gamma + 2 \ln \left( \frac{\mu R}{2} 
\right) \right),$$ and take the limit $|\lambda_B| \to 0$.  
In other words the usual (i.e. $\delta$ function) description of contact 
interactions is indeed equivalent to the appropriate $|\nu| \rightarrow 0$ 
limit of the Schroedinger equation \eqref{sedf} subject to the boundary 
conditions \eqref{pbc}. This suggests that the boundary conditions 
\eqref{pbc} do indeed capture the effect of contact interactions at general 
$\lambda_B$. This has been argued to be true in
\cite{AmelinoCamelia:1994we,Kim:1996rz}. 

\subsection{Derivation of the scattering amplitude}\label{abap}

In this section we will derive the scattering amplitude for the 
Schroedinger equation \eqref{sedf} subject to the boundary conditions 
\eqref{pbc}. We assume $|\nu|<1$.

We wish to find scattering state solutions at energy 
$E= \frac{k^2}{2m}$ of the Schroedinger equation for this particle; 
i.e. $k$ is the magnitude of the momentum of the particle incident on the 
solenoid. The most general solution of the Schroedinger equation that meets the boundary conditions for $\psi_m(r)$  
at the origin ($m \neq 0$) is 
\begin{equation}\label{pwe}
\psi({\vec r})= \sum_{n >0} a_{n} e^{i n \theta}  J_{n + \nu}(k r) 
+ \sum_{n>0} a_{-n} e^{-in \theta} J_{n-\nu}(k r) + a_0 J_{|\nu|}(kr) + b_0 J_{-|\nu|}(kr).
\end{equation}
The scattering solution we wish to find obeys the boundary condition 
\eqref{pbc}; moreover at large $r$ its ingoing piece (part proportional to 
$e^{-ikr}$) must reduce to that of the incoming wave $e^{ikx}$. It is 
not difficult to see that the unique solution  
that meets our boundary conditions is given by (see Appendix C of \cite{Jain:2014nza} for the detailed derivation for the special case $w=0$)
\begin{equation}\label{scatwf} 
\begin{split}
 \psi({\vec r})=& \sum_{n=1}^\infty i^n e^{-i \frac{\pi \nu }{2} } J_{n +\nu}(kr) e^{i n \theta}
+ 
\sum_{n=1}^\infty i^{n} e^{i \frac{\pi \nu }{2} } J_{n -\nu}(kr) e^{-i n \theta} \\
+ & \frac{ \Gamma(|\nu| +1) \left( \frac{2}{k} \right)^{|\nu|}
J_{|\nu|}(kr) + w R^{2 |\nu|} \Gamma(1-|\nu|) \left( \frac{k}{2} \right)^{|\nu|}
J_{-|\nu|}(kr)}{\Gamma(|\nu| +1) \left( \frac{2}{k} \right)^{|\nu|}
e^{i \frac{ \pi |\nu|}{2}} + w R^{2 |\nu|} \Gamma(1-|\nu|) \left( \frac{k}{2} \right)^{|\nu|}
e^{-i \frac{ \pi |\nu|}{2}}}.
\end{split}
\end{equation}

 At large $r$, $\psi({\vec r})$ reduces to 
\begin{equation*}  \frac{1}{\sqrt{ 2 \pi k r} } \left( e^{i \frac{\pi}{4} }  \delta(\theta - \pi) e^{-ikr} 
+ H(\theta) e^{-i \frac{\pi}{4} }  e^{ikr} \right),
\end{equation*}
where,
\begin{equation}\label{heq}
\begin{split}
H(\theta)= &\sum_{n=1}^\infty \left( e^{-i \pi \nu} e^{i n \theta} 
+ e^{i \pi \nu} e^{- i n \theta} \right)\\
&+  \frac{ \Gamma(|\nu| +1) \left( \frac{2}{k} \right)^{|\nu|} 
e^{- i \frac{\pi |\nu|}{2}} + w R^{2 |\nu|} \Gamma(1-|\nu|) \left( \frac{k}{2} \right)^{|\nu|}  
e^{i \frac{ \pi |\nu|}{2}} } 
{\Gamma(|\nu| +1) \left( \frac{2}{k} \right)^{|\nu|}
e^{i \frac{ \pi |\nu|}{2}} + w R^{2 |\nu|} \Gamma(1-|\nu|) \left( \frac{k}{2} \right)^{|\nu|}
e^{-i \frac{ \pi |\nu|}{2}}}.
\end{split}
\end{equation}
Now we can write 
\begin{equation}\label{hee}\begin{split} 
\sum_{n=1}^\infty \left( e^{-i \pi \nu} e^{i n \theta} 
+ e^{i \pi \nu} e^{- i n \theta} \right)&= \left( \sum_{n=1}^\infty 2 \cos (\pi \nu) \cos (n \theta)  \right)
+ \left( \sum_{n=1}^\infty  2 \sin (\pi \nu) \sin (n\theta) \right) \\
&= \left(  \cos (\pi\nu)  + \sum_{n=1}^\infty 2 \cos (\pi \nu) \cos (n \theta)  \right)  -\cos(\pi\nu)
+ \left( \sum_{n=1}^\infty  2 \sin (\pi \nu) \sin (n\theta) \right)\\
&= 2 \pi \cos (\pi \nu) \delta(\theta) -\cos(\pi\nu)
+ \left( \sum_{n=1}^\infty  2  \sin (\pi \nu) \sin (n\theta)  \right)\\
&= 2 \pi \cos (\pi \nu) \delta(\theta) + \sin (\pi \nu) {\rm Pv}
\left(\cot \left(\frac{\theta}{2}\right) \right) -\cos(\pi\nu).\\
\end{split}
\end{equation}
Substituting in \eqref{heq}
\begin{equation}
\begin{split}
H(\theta) &= 2 \pi \cos (\pi \nu) \delta(\theta) + \sin (\pi \nu) {\rm Pv}
\left(\cot \left(\frac{\theta}{2}\right) \right)\\&+\frac{ \Gamma(|\nu| +1) \left( \frac{2}{k} \right)^{|\nu|} 
e^{- i \frac{\pi |\nu|}{2}} + w R^{2 |\nu|} \Gamma(1-|\nu|) \left( \frac{k}{2} \right)^{|\nu|}  
e^{i \frac{ \pi |\nu|}{2}} } 
{\Gamma(|\nu| +1) \left( \frac{2}{k} \right)^{|\nu|}
e^{i \frac{ \pi |\nu|}{2}} + w R^{2 |\nu|} \Gamma(1-|\nu|) \left( \frac{k}{2} \right)^{|\nu|}
e^{-i \frac{ \pi |\nu|}{2}}} -\cos(\pi\nu)\\
&= 2 \pi \cos (\pi \nu) \delta(\theta) + \sin (\pi \nu) {\rm Pv}
\left(\cot \left(\frac{\theta}{2}\right) \right)\\&-i\sin(\pi|\nu|)\frac{ \Gamma(|\nu| +1) \left( \frac{2}{k} \right)^{|\nu|}e^{i \pi |\nu|} 
- w R^{2 |\nu|} \Gamma(1-|\nu|) \left( \frac{k}{2} \right)^{|\nu|}  
 } 
{\Gamma(|\nu| +1) \left( \frac{2}{k} \right)^{|\nu|}
e^{i \pi |\nu|} + w R^{2 |\nu|} \Gamma(1-|\nu|) \left( \frac{k}{2} \right)^{|\nu|}
}.
\end{split}
\end{equation}

In order to compute the scattering amplitude, we must rewrite the 
wave function as a plane wave plus a scattered piece; at large $r$ 
\begin{equation} \label{conwe}
\psi(r) = e^{ikx} + \frac{h(\theta) e^{-i \frac{\pi}{4} }  e^{ikr}}{\sqrt{ 2 \pi k r} } .
\end{equation}
We find 
\begin{equation}\label{hto}
h(\theta)= H(\theta) - 2 \pi \delta(\theta),
\end{equation}
so that 
\begin{equation}\label{htff}
\begin{split}
h(\theta)&= 2 \pi  \left( \cos (\pi\nu) - 1 \right) \delta(\theta) + \sin (\pi \nu) {\rm Pv}  
\left( \cot \left(\frac{\theta}{2}\right) \right)\\&-i\sin(\pi|\nu|)\frac{ \Gamma(|\nu| +1) \left( \frac{2}{k} \right)^{|\nu|}e^{i \pi |\nu|} 
- w R^{2 |\nu|} \Gamma(1-|\nu|) \left( \frac{k}{2} \right)^{|\nu|}  
 } 
{\Gamma(|\nu| +1) \left( \frac{2}{k} \right)^{|\nu|}
e^{i \pi |\nu|} + w R^{2 |\nu|} \Gamma(1-|\nu|) \left( \frac{k}{2} \right)^{|\nu|}
}.
\end{split}
\end{equation}
This yields \eqref{scatsol}.

\bibliographystyle{JHEP}
\bibliography{bsJul4}
\end{document}